\documentclass[aps,prl,twocolumn,groupedaddress,showpacs]{revtex4}
\usepackage[latin1]{inputenc}
\usepackage[english]{babel}
\usepackage{setspace}  
\usepackage{graphicx}
\usepackage{amssymb}
\usepackage{amscd}
\usepackage{rotating}
\usepackage{color}
\usepackage{epstopdf}

\begin{document}

\title{Discerning Incompressible and Compressible Phases of Cold Atoms in Optical Lattices}

\author{V.W. Scarola$^{1,2}$, L. Pollet$^{2}$, J. Oitmaa$^3$, M. Troyer$^2$}
\affiliation{$^1$Department of Chemistry and Pitzer Center for Theoretical Chemistry, University of California, Berkeley, California 94720, USA\\
$^2$Theoretische Physik, ETH Zurich, 8093 Zurich, Switzerland \\
$^3$School of Physics, The University of New South Wales, Sydney, New South Wales 2052, Australia}

\date{\today}

\begin{abstract}
Experiments with cold atoms trapped in optical lattices offer the potential to realize a variety of novel phases but suffer from severe spatial inhomogeneity that can obscure signatures of new phases of matter and phase boundaries.   We use a high temperature series expansion to show that compressibility in the core of a trapped Fermi-Hubbard system is related to measurements of changes in double occupancy.   This {\em core compressibility} filters out edge effects, offering a direct probe of compressibility independent of inhomogeneity.  A comparison with experiments is made.
\end{abstract}
\maketitle

The search for stable quantum many-body phases forms the basis of quantum condensed matter, quantum chemistry and elementary particle physics.  Stable states often arise as a consequence of energy gaps that set an energy scale for resilience.   Examples of gapped condensed matter phases 
include superconductors and Mott insulators.  In these phases gaps arise in the presence of (or as a result of) interactions to form many-body states that resist small perturbations from the environment.  Gaps play a role in spectacular quantum effects in the solid state including zero resistance in superconductors or a dramatic rise in resistance in Mott insulators.  The high $T_c$ compounds are believed to harbor both states in a possible union \cite{anderson1997} that leads to an enhancement of the superconducting energy gap and therefore $T_c$ in a doped correlated Mott insulator.  
Electronic models of the high $T_c$ compounds (repulsive two-dimensional Fermi-Hubbard models) are believed to have rich phase diagrams and may even exhibit this unique type of superconductivity.  Yet the low temperature solution of the two-dimensional Hubbard model remains elusive.  Recent work \cite{hofstetter2002,trebst2006} seeks to emulate the Hubbard models in an ideal setting, with cold atoms in optical lattices, in a search for a superconducting state and other phases.  

Optical trapping and cooling of cold atomic gasses allows the preparation of nearly ideal manifestations of Bose and Fermi-Hubbard models  in the laboratory \cite{jaksch1998,greiner2002,jordens2008,bloch2008}.  A broad variety of many-body phases have been predicted \cite{lewenstein2007}.  However, few techniques exist for experimentally observing new phases and their properties.  Techniques currently in use include time of flight imaging of the momentum distribution (predominately in bosonic systems) and noise correlations \cite{noise}, optical molecular spectroscopy of pair correlation functions \cite{partridge2005} and  Bragg spectroscopy \cite{stoferle2004}.  Proposed techniques include using edge currents in trapped rotating lattices \cite{scarola2007} or Fourier sampling of time-of-flight images to reveal new correlation functions \cite{duan2006}.  

 However, harmonic trapping potentials, inherent in most experiments, present a major difficulty in realizing and observing bulk phases of the Hubbard model with optical lattice experiments: e.g., a large portion of the sample becomes compressible near the edges even when a Mott insulator has formed in the center. Extracting the physics of a homogeneous bulk system by 
separating the approximately homogeneous behavior in the flat center of the trap from the surrounding inhomogeneous system is an intrinsically difficult challenge in these experiments and for comparison to theory
\cite{batrouni2002,pollet2004,wessel2004,scarola2006,campo2007,helmes2008,roscilde2008}.

In this Letter we use a high temperature series expansion \cite{henderson1992,haaf1992,oitmaa2006} to directly relate recent experiments \cite{jordens2008} to (a measure of) ``core compressibility'', the compressibility of the atoms near the center of the sample, excluding the edges.  We argue that this measure of the core compressibility can  be used to detect stable, incompressible phases and offers a valuable tool for mapping out phase diagrams.  
 
We begin our analysis with the Fermi-Hubbard model of cold fermionic atoms in optical lattices for equal pseudo-spin populations where the spin index refers to different hyperfine levels:
\begin{eqnarray}
  H=-  t\sum_{\langle i,j \rangle, \sigma} ( c^{\dagger}_{\sigma,i} c^{\phantom{\dagger}}_{\sigma,j} + \text{h.c.})+U\sum_{i} n_i^{\uparrow}n_i^{\downarrow}
  -\sum_{i}\mu_i n_i.
 \label{Hubbard}
\end{eqnarray}
Here the site dependent chemical potential, $\mu_i=\mu-\gamma {\bf R}_i^2$ varies with the discrete spatial coordinates, ${\bf R}_i=(i_x,i_y,i_z)$, as a result of the parabolically confining laser beam waist and magnetic trapping potentials.  
The $s$-wave interaction yields an on-site repulsive interaction, of strength $U$, which is tunable via a Feshbach resonance.  The 
hopping between nearest neighbor sites, $t$, changes with the lattice depth and is therefore tunable with the laser intensity.  The number operator 
 $n_i=n_i^{\uparrow}+n_i^{\downarrow}$ measures the number of fermions at a site $i$.

For our quantitative calculations we use parameters relevant to recent experiments.  Ref.~\cite{jordens2008}  places two hyperfine species of $^{40}$K in a simple cubic optical lattice with low hopping $t=0.05-0.2$ kHz, a tunable on-site term, $U=0-9$kHz and $\gamma= 0.003-0.005$ kHz. Here and in the following we work in kHz by setting $h=1$ and $k_{\text{B}}=1$.  We also set $\gamma=0.00384$ kHz, unless otherwise noted, and choose all other parameters to make contact with Ref.~\cite{jordens2008}.  Suitably chosen chemical potentials, $\mu\sim 1-7$ kHz,  yield total particle numbers, $N\sim10^4-10^6$.  Estimates in Ref.~\cite{jordens2008} find that the largest unknown, lattice temperature, can be kept below $U$ with values as low as $T\approx 0.8$ kHz and possibly lower.

To theoretically analyze observable signatures of incompressibility in a trapped Hubbard model we use a high temperature series expansion of the grand partition function $\mathcal{Z}=\text{Tr} \exp(-\beta H)$ about the atomic, $t=0$, limit.
Such high temperature series expansions have a long history \cite{oitmaa2006} and yield exact results for thermodynamic quantities of the Hubbard model in the limit $\beta t \ll 1$.  Note that all experiments done with fermions in optical lattices currently lie in this high temperature regime when parameters are tuned to $U\gg t$.  The high temperature series therefore offers a quantitatively reliable tool to compare to experiments.  Our approach complements recent dynamical mean field studies that can be applied to lower temperatures \cite{leo2008}.  We use up to $10^{\text{th}}$ order in the expansion of the grand potential \cite{henderson1992},  $\Omega = - \ln{(\mathcal{Z})}/\beta $, to extract thermodynamic quantities for a uniform system  ($\gamma=0$). In terms of the series coefficients, $X^{(m)}$, the expansion reads:  
\begin{equation}
-\beta \tilde{\Omega}=\ln z_{0}+\sum_{m=2}^{\infty} (\beta t/z_{0})^{m}X^{(m)}(w,\zeta ),
\end{equation}
where  $\zeta = \exp(\beta\mu)$ is the fugacity of the uniform system, $ \tilde{\Omega}\equiv\Omega/N$, $w=\exp(-\beta U)$ and $ z_{0}=1+2\zeta+\zeta^2w$ is the partition function of a single site in the atomic limit.  In a local density approximation (LDA) we assume that each site of the trapped system can be approximated with parameters for a uniform system.  With the replacement $\zeta\rightarrow x_i=\exp(\beta \mu_i)$ the LDA becomes $\Omega^{\text{LDA}}=\sum_i\Omega_i$.  

To show that the LDA  is an excellent approximation for the high temperature regime studied here we also compute the exact second order contribution to the grand potential in a trapped system:
\begin{equation}
-\beta\Omega=\sum_i\ln z_{0,i} +(t\beta)^2\sum_{i , j\in n.n.}\frac{\tilde{X}_{ij}^{(2)}}{z_{0,i}z_{0,j}}+\mathcal{O}((\beta t)^4),
\end{equation}
where the second sum double counts nearest neighbors and \begin{eqnarray}
\tilde{X}^{(2)}_{ij}&=& 
I_{\delta,-\delta} [x_i+x_i^2x_jw]
+I_{-\delta,\delta}  [x_j+x_j^2x_iw] \nonumber \\
&+&x_ix_j[I_{-\delta-U,\delta+U}
+I_{\delta-U,-\delta+U}]  \nonumber \\
&+&x_i^2w I_{\delta+U,-\delta-U} +x_j^2w I_{-\delta+U,\delta-U}. 
\end{eqnarray}
In the above expression the quantities $\delta\equiv\mu_i-\mu_j$ and 
$
I_{\Delta,-\Delta}\equiv
(\exp(\beta\Delta)-1-\beta\Delta)/(\beta\Delta)^2
$
simplify to the uniform limit for $\delta =0$ and $I_{0,0}=1/2$.  The LDA is recovered in the limit 
$\tilde{X}_{ij}\rightarrow X_r^{\text{LDA}}\equiv2(1-w)x_r^2/\beta U+x_r(1+x_r^2w)$
yielding a simple expression for the grand potential of a trapped system in the continuum approximation (valid for large particle numbers):
$-\beta\Omega^{\text{LDA}}\approx  \int d^{d}r [\ln z_{0,r} +z(t\beta/ z_{0,r})^2 X_r^{\text{LDA}} ]$, 
where $z$ is the coordination number.  This second order high temperature expansion holds for any bipartite lattice in any dimension.
As we show below there are no discernible differences between the LDA and the exact second order results for the parameters studied here.  
      
\begin{figure}[t] 
   \centering
   \includegraphics[width=3in]{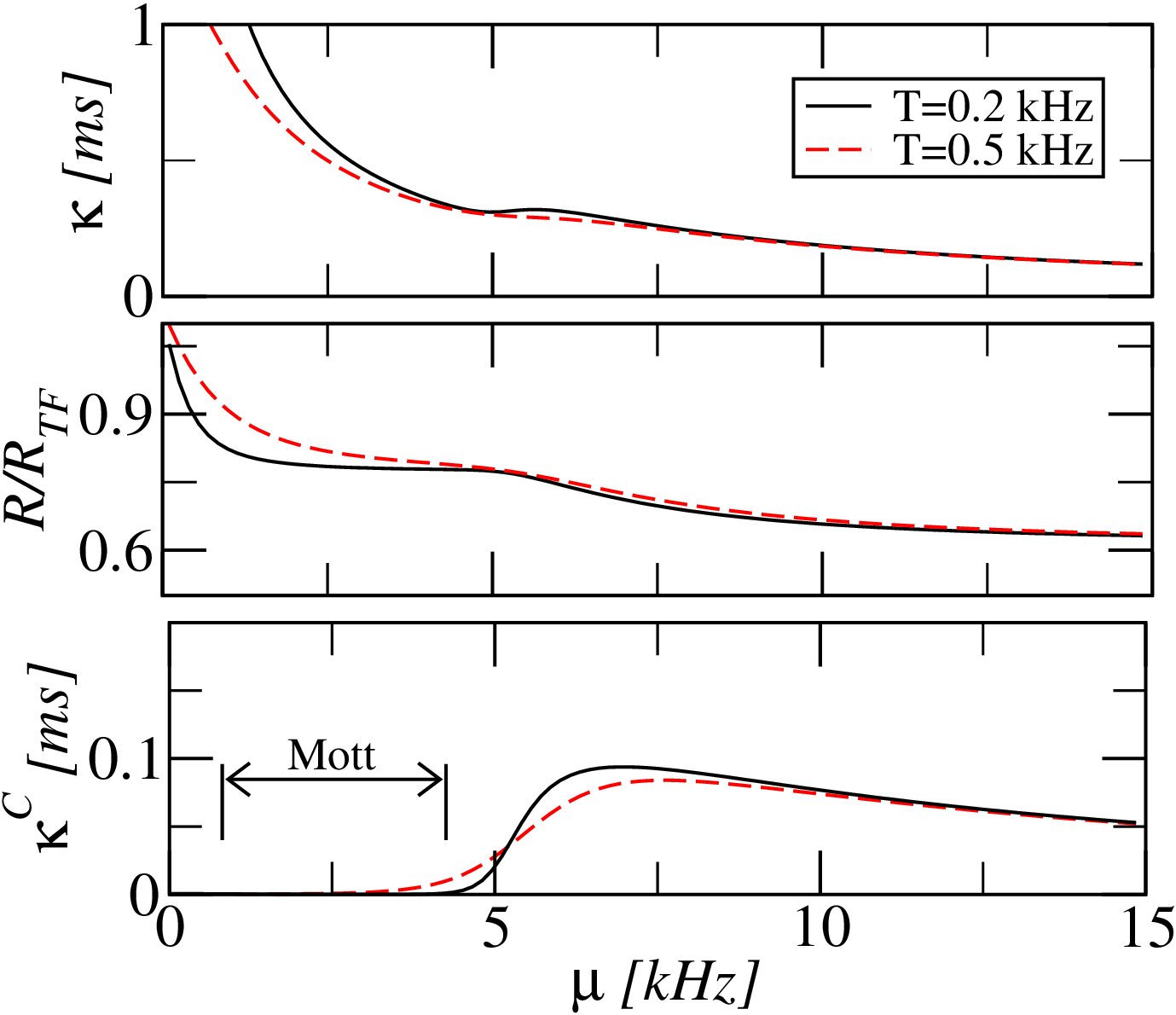} 
   \caption{Total compressibility (top panel), system radius (middle panel) and the core compressibility (bottom panel) plotted as a function of the chemical potential in the trap center for $t=0.054$ kHz  and $U=5$ kHz.  The arrows in the bottom panel indicate the Mott regime at $T= 0.2$ kHz. }
   \label{fig1}
\end{figure}

We first focus on the compressibility which can distinguish the incompressible ground state of a gapped phase from a compressible metallic phase in a homogeneous system.
  The compressibility per particle is  defined  as 
$
\kappa = N^{-1}\sum_i \partial n_i / \partial \mu
$
 where $N=\sum_i \langle n_i \rangle=\sum_i x_i \partial_{x_i} (-\beta \tilde{\Omega})$.   
 We compute the compressibility with the $10^{\text{th}}$ order series but find essentially no distinction from the second order results for $\beta t\lesssim 0.9$.  The top panel of Fig.~\ref{fig1} plots the compressibility as a function of the chemical potential in the center of the trap, $\mu$, for two different values of $T$.    
  At low chemical potentials ($-1.0$ kHz $\lesssim \mu \lesssim 0.8 $ kHz) the system forms a compressible Fermi-lattice gas.  At higher chemical potentials   ($0.8$  kHz $ \lesssim  \mu \lesssim 4.2 $ kHz)  the system forms a Mott insulator at the trap center.  Here and in what follows we define the Mott regime with a central density  $\langle n_i \rangle =1\pm 0.01$.  A weak, barely distinguishable feature around $\mu = 5.5$ kHz in the top panel indicates that a finite fraction of the system has become incompressible.  Otherwise, we find that $\kappa$ is a nearly smooth function of all experimentally relevant parameters:  the total compressibility incorporates the edges of the system which overwhelms signatures of the Mott transition.   

We next study the system radius, $R$, because the compressibility is not directly measurable in experiments.  $R$ is defined as the root mean square of the distance averaged with respect to the density \cite{schneider2008}.  The central panel of Fig.~\ref{fig1} plots $R$ in units of the Thomas-Fermi radius, $R_{\text{TF}}=a(3N/4\pi)^{1/3}$, where $a$ is the lattice spacing.  The plateau in the middle panel of Fig.~\ref{fig1} results from a combination of edge effects and incompressibility at the trap center.  As the chemical potential increases, we add more particles to the edge gas, and the size $R$ scales as the size $R_{\rm TF}$, the relevant length scale for the edges.  A similar plateau can be seen in the compressibility in units of the Thomas-Fermi compressibility at a value $\kappa / \kappa_{\rm TF} = 1$.  Nonetheless, we have again found only weak, edge dependent features that indicate the formation of a Mott insulator.   

\begin{figure}[t] 
   \centering
   \includegraphics[width=3in,angle=0]{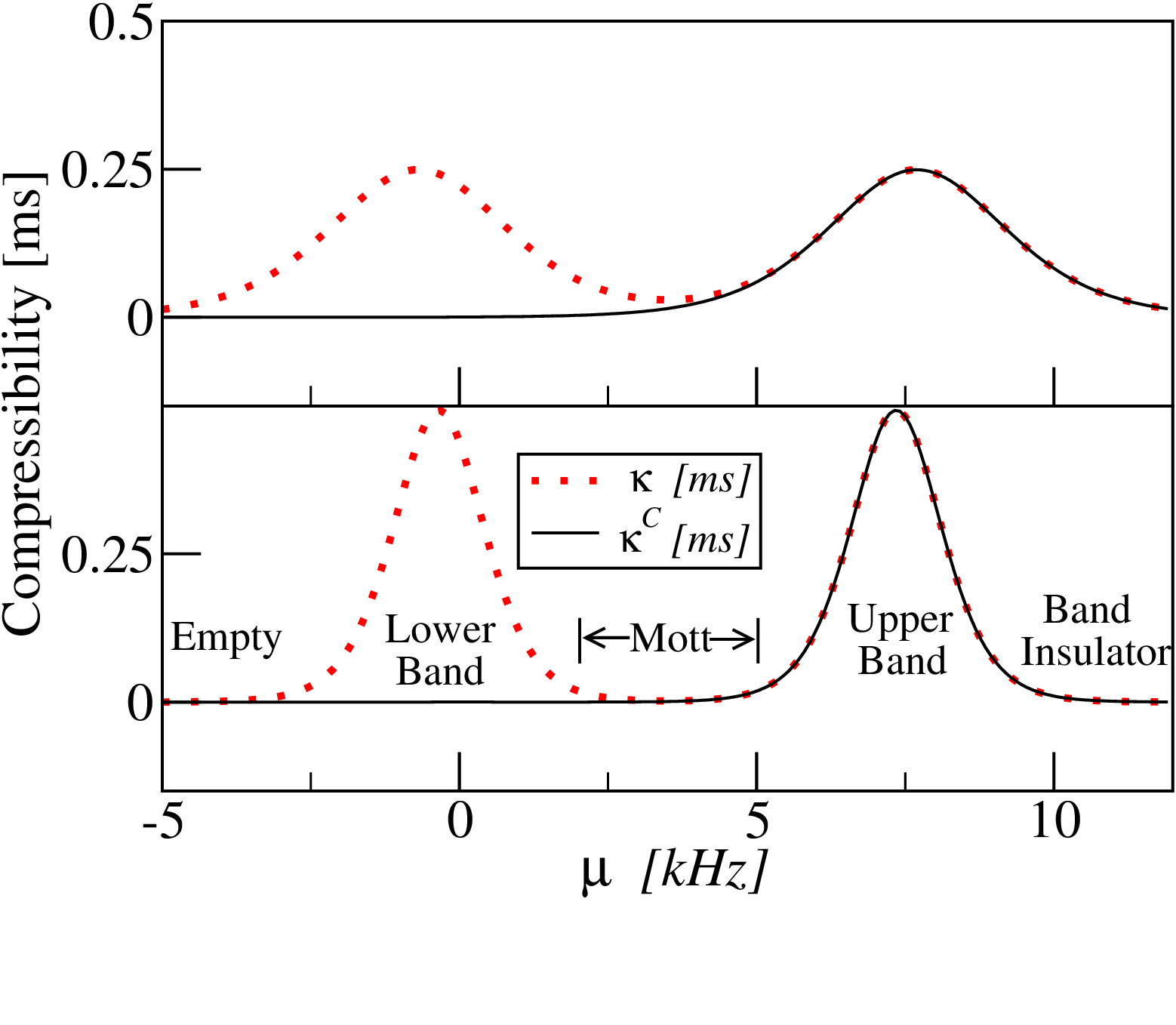} 
   \caption{ Compressibility (dotted line) and core compressibility (solid line) versus chemical potential in a uniform system, $\gamma=0$, with $U=7$ kHz and $t=0.054$ kHz.  The top (bottom) panel sets $T=1$ kHz ($T=0.5$ kHz).  The three incompressible regions correspond to a pinning of the density at 0, 1 and 2 for chemical potentials near $\mu=-U/2,U/2$ and $3U/2$, respectively.  For $\mu\lesssim U/2$,  $\kappa^{\text{C}}$ vanishes because the system shows very little double occupancy.     
   }
   \label{fig2}
\end{figure}

We are thus looking for a robust technique that probes the gapped phase without resorting to edge effects. We will show that the  $\emph{core compressibility}$ per particle fulfills this requirement:
\begin{equation}
\kappa^{\text{C}} \equiv  N^{-1}\partial D /\partial \mu,
\end{equation} 
where the double occupancy is conventionally defined as
$
D=  \partial (\Omega) /\partial U=\sum_{i} \langle n_i^{\uparrow}n_i^{\downarrow} \rangle.
$
Our definition of $D$ differs from the definition in Ref.~\cite{jordens2008}.  In Ref.~\cite{jordens2008} a quantity related to its dimensionless derivative, $\partial D/ \partial N$, was measured, which directly relates to the ratio of core and total compressibility  (without knowledge of $\mu$) via
$
\partial D / \partial N = ( \partial D / \partial \mu ) ( \partial \mu / \partial N) = \kappa^{\text C} / \kappa.
$

\begin{figure}
   \centering
   \includegraphics[width=3in]{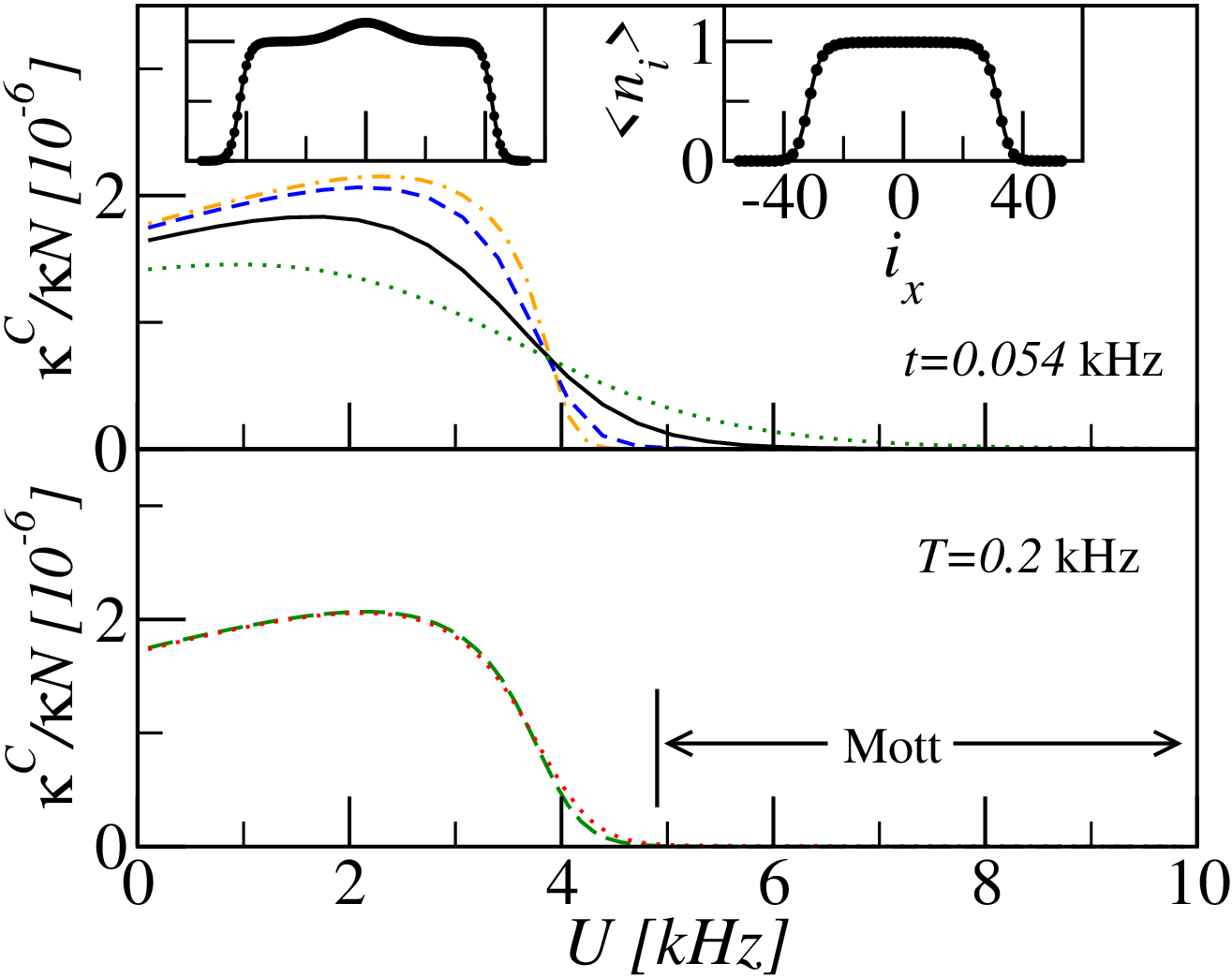} 
   \caption{ 
   The core compressibility ratio versus on-site interaction strength for $\mu=4$ kHz.  The top panel fixes $t=0.054$ kHz for several temperatures, 
   $T= 0.07$ kHz (dot-dashed), 0.2 (dashed), 0.5 (solid) and 1.0 (dotted).   
    The  bottom panel fixes $T=0.2$ kHz for $t=0.01$ kHz (dashed) and 0.1 (dotted). 
   The insets plot the density along a cross section in the cubic lattice for the same parameters as the bottom panel of Fig.~\ref{fig2} but in a trap and with $\mu=6.5$ kHz (left) and  $\mu=3.5$ kHz (right).   The circles are computed in a full second order expansion for the density while the solid line is computed in the second order LDA.  The Mott gap pins the density near $\langle n_i \rangle=1$ at the trap center (right panel).  For  $\mu>U/2$ (left panel) a central compressible region arises with $n_i>1$ at the trap center.}
   \label{fig3}
\end{figure}

In Fig.~\ref{fig2} we plot $\kappa^{\text{C}}$ and $\kappa$ versus $\mu$ for two different temperatures in a uniform system and find that $\kappa^{\text{C}}$ is essentially identical  
to $\kappa$ when the system has doubly occupied sites, but is zero otherwise. This is a key feature with useful implications for trapped systems:  
$\kappa^{\text{C}}$, by taking the derivative of the double occupancy, measures the compressibility of the core region with density larger than one (see the insets of Fig.~\ref{fig3}) and is therefore insensitive to the edges. 

A comparison between the top and bottom panels of Fig.~\ref{fig2} shows that $\kappa^{\text{C}}$ and $\kappa$ agree at low temperatures and high chemical potentials.  This can be understood by considering a single site in the atomic limit: 
$
(\kappa^{\text{C}}/\kappa)_{t=0}=(1+\zeta)[2+\zeta+w/\zeta]^{-1}.
$
At zero temperature we can see the implicit cutoff in the chemical potential $(\kappa^{\text{C}}/\kappa)_{t=0}\rightarrow\theta(\mu-U/2)$.  Measurements of $\partial D / \partial N$ therefore exclude low chemical potentials (and thus edge effects) in trapped systems, yielding a measure of the core compressibility ratio (CCR),  $\kappa^{\text{C}}/\kappa$, at low temperatures, $T\ll U$.  

The LDA  is an excellent approximation for the parameters considered here.  The circles in the insets of 
Fig.~\ref{fig3} plot the density computed without the LDA.  Comparison with the LDA results (solid line) shows remarkable agreement for $\beta t \lesssim 1$.  
We have also compared the compressibility computed with and without the LDA and have found very little distinction (less than $4\%$) 
in the regime of validity of our high temperature series  $\beta t \lesssim 1$.  The LDA is thus a good approximation and the 
high temperature series expansion yields a highly accurate tool for quantitative comparison with ongoing experiments with 
strongly interacting ($t/U \ll 1$) fermions in optical lattices.  

\begin{figure}[t] 
   \centering
   \includegraphics[width=3in]{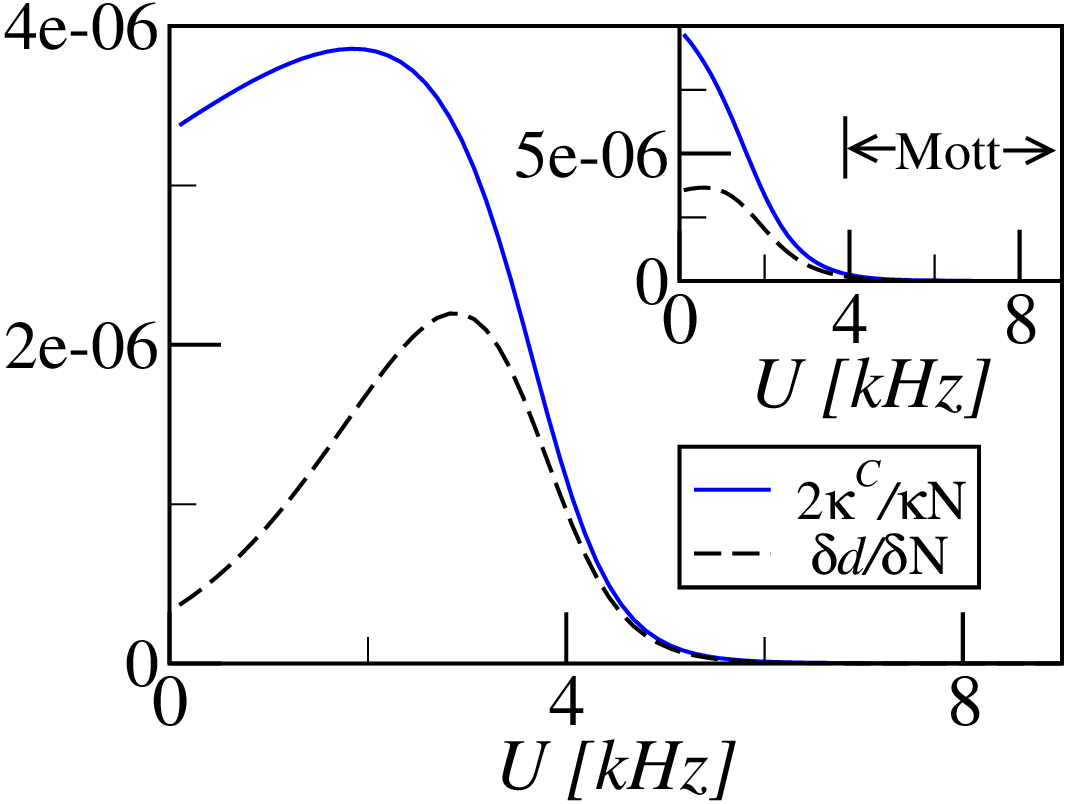} 
   \caption{
    The core compressibility ratio (solid line) and the experimentally observed quantity $\delta d/\delta N$ (dashed line), plotted as a function of the on-site interaction strength for $\mu=4$ kHz,  $t=0.054$ kHz and $T=0.3$ kHz. The central density is unity for $U \gtrsim 6$ kHz.  The two lines merge when there is little double occupancy.  The inset plots the same but the values are obtained by matching the entropy of  $N=80000$ atoms in the dipole trap (at $T_D/T_F=0.15$) to those in the lattice with $t=0.054$.}
\label{fig4}
\end{figure}

We now make contact with recent experiments of  Ref.~\cite{jordens2008}, reporting a measurement of the quantity $\delta d/\delta N$ where $\delta$ stands for an approximate derivative taken by linear fitting to experimental data.  This 
experiment measures the double occupancy fraction $d\equiv 2D/N$ instead of the double occupancy $D$.  Estimating  
 $\delta d/\delta N \approx \partial d/\partial N= (2\kappa^{\text{C}}/\kappa N)-D/N^2$ we find that the difference is small near unity filling implying 
that Ref.~\cite{jordens2008} measures the CCR.  Note that $\delta N$ must be carefully chosen to ensure the proper phase \cite{Bloch_private}.  Fig.~\ref{fig4} compares $2\kappa^{\text{C}}/\kappa N$ and  $\delta d/\delta N$, showing agreement near the Mott transition with only small differences of order $D/N^2$.  However, 
we assert that a measurement of the quantity $\delta D /\delta N\approx \kappa^{\text{C}}/\kappa $ provides a better and more direct connection to the core compressibility than $\delta d/\delta N$. 

Both the CCR and  $\delta d/\delta N$ plotted in Fig.~\ref{fig4} show a distinct signature of the transition to the incompressible Mott phase.  At low $U$ the core of the 
system lies in the compressible regime of the Fermi-Hubbard phase diagram, $n_i  >1$.  The peak structure in Fig.~\ref{fig4} originates from the choice of $\mu$ and the peaks in Fig.~\ref{fig2}.  Upon increasing $U$ we enter the Mott regime,  $n_i  =1$.  (Note that compressibility alone cannot distinguish the Mott phase from a weakly compressible metallic phase.)  Here the center of the trapped system opens a gap and the core compressibility drops exponentially to zero.  The zeroing of the core compressibility (and the CCR) is therefore an indicator for the onset of a Mott insulating phase in the sample center.  Recall that the total compressibility $\kappa$ shows very little structure as we enter the Mott phase of a trapped system, and the signal thus originates from the change in core compressibility (compare the top and bottom panels of Fig.~\ref{fig1}).  

The inset of Fig.~\ref{fig4} makes a more direct comparison with Ref.~\cite{jordens2008}.  Here we match entropy in the dipole trap at a temperature 
$T_D$ with the entropy in the lattice at fixed $N$ to obtain a lattice temperature $T$.  The inset is plotted for the coldest temperatures ascertained in experiment.  For larger temperatures, $T_D/T_F=0.195$, where $T_F$ is the Fermi temperature in the dipole trap, the Mott phase is less robust.  

We also find that the precipitous drop in core compressibility remains robust over a wide parameter range.   The top panel of Fig.~\ref{fig3} plots $\kappa^{\text{C}}/\kappa N$  for several different temperatures.  We find that the CCR shows a clear drop upon entering the Mott phase for temperatures 
well below $U$.  For temperatures comparable to the Mott gap, the center of the trapped system becomes compressible 
and the clear signature of a Mott phase vanishes (dotted line).  A temperature fixed point appears for $U\approx\mu$ such that $\langle H \rangle \approx 0$.   Here one finds a crossover from a high temperature Fermi-gas to a Mott phase.  The bottom panel of Fig.~\ref{fig3} varies the hopping to demonstrate that there is only a small shift with different hoppings.

We have shown that observations of double occupancy of cold atoms in optical lattices reveal the core compressibility in a trapped Fermi-Hubbard model.  This core compressibility clearly indicates the onset of incompressible states and describes recent measurements that show evidence for the Mott transition \cite{jordens2008}.  Ongoing work will generalize our proposed technique to bosonic systems. The core compressibility implicitly excludes edge effects to reveal compressibility near the trap center.  Transitions to incompressible phases (e.g., metal-insulator transitions) nucleated at the sample center can be readily identified in experiments and compared with compressibility computed in uniform Hubbard models.  

We thank T. Esslinger and his group, F. Hassler and S. Huber for valuable 
discussions.  LP, MT and VS thank the Swiss National Science Foundation for support.
%%%%%%%%%%%%%%%%%%%%%%%%%%%%%%%%%%%%%%%%%%%%
%%%%%%%%%%%%%%%%%%%%%%%%%%%%%%%%%%%%%%%%%%%%

\end{document}